\documentclass[12pt]{article}
\usepackage{amsmath,amssymb,bm,epsfig,afterpage}
\usepackage{cite}

\usepackage{color}

\renewcommand{\thefootnote}{\fnsymbol{footnote}}
\newcommand{\vev}[1]{{\langle{#1}\rangle}}
\newcommand{\MEW}{m_Z}

\begin{document}
\title{
\begin{flushright}
\begin{minipage}{0.2\linewidth}
\normalsize
WU-HEP-15-10 \\*[50pt]
\end{minipage}
\end{flushright}
{\Large \bf 
LHC phenomenology of natural MSSM with non-universal gaugino masses at the unification scale 
\\*[20pt]}}
\author{Hiroyuki~Abe$^a$,\footnote{
E-mail address: abe@waseda.jp} \, \ 
Junichiro~Kawamura$^a$\footnote{
E-mail address: junichiro-k@ruri.waseda.jp} \ and
Yuji~Omura$^b$\footnote{
E-mail address: yujiomur@eken.phys.nagoya-u.ac.jp}\\*[20pt]
{\it \normalsize 
$^a$Department of Physics, Waseda University, 
Tokyo 169-8555, Japan} \\
{\it \normalsize 
$^b$Kobayashi-Maskawa Institute for the Origin of Particles and the Universe (KMI), } \\
{\it \normalsize
Nagoya University, Nagoya 464-8602, Japan} \\*[50pt]
}
\date{
\centerline{\small \bf Abstract}
\begin{minipage}{0.9\linewidth}
\medskip 
\medskip 
\small
In this letter, we study collider phenomenology in the supersymmetric Standard Model 
with a certain type of non-universal gaugino masses at the gauge coupling unification scale, 
motivated by the little hierarchy problem. 
In this scenario, especially the wino mass is relatively large compared to the gluino mass at the  
unification scale, and the heavy wino
can relax the fine-tuning of the higgsino mass parameter, so-called $\mu$-parameter. 
Besides, it will enhance the lightest Higgs boson mass due to the relatively large left-right mixing of
top squarks through the renormalization group (RG) effect.
Then $125$ GeV Higgs boson could be accomplished, 
even if the top squarks are lighter than $1$ TeV and the $\mu$ parameter is within a few hundreds GeV. 
The right-handed top squark tends to be lighter than the other sfermions due to the RG runnings, 
then we focus on the top squark search at the LHC.
Since the top squark is almost right-handed and the higgsinos are nearly degenerate, 
$2b + E_T^{\rm miss}$ channel is the most sensitive to this scenario.  
We figure out current and expected experimental bounds 
on the lightest top squark mass and model parameters at the gauge coupling unification scale.
\end{minipage}
}

\begin{titlepage}
\maketitle
\thispagestyle{empty}
\clearpage
\tableofcontents
\thispagestyle{empty}
\end{titlepage}

\renewcommand{\thefootnote}{\arabic{footnote}}
\setcounter{footnote}{0}

\section{Introduction}

The Minimal Supersymmetric Standard Model (MSSM) 
is a good candidate for new physics to be found at the LHC~\cite{Martin:1997ns}.
One of the strong motivations to consider a low-scale supersymmetry (SUSY) is 
to ensure the stability of the electroweak (EW) scale against huge radiative corrections.
The other incentives are to provide suitable candidates for dark matters 
and its prediction of the gauge coupling unification 
around $2\times 10^{16}\ {\rm GeV}$, 
which indicates the existence of the Grand Unification Theories (GUTs).

The natural explanation of the EW scale requires light supersymmetric particles (sparticles), but there has been no signature of SUSY at the LHC Run-I. 
Moreover, the observed value of the Standard-Model-like Higgs boson mass might imply heavy sparticles 
since it requires heavy top squark masses to enhance the Higgs boson mass up to
$125$ GeV~\cite{Aad:2012tfa,Chatrchyan:2012ufa}, by the radiative corrections.
Such heavy sparticles bring a fine-tuning problem~\cite{Barbieri:1987fn} 
that can be seen in the minimization condition of the Higgs potential, 
\begin{eqnarray}
m_Z^2 
         &\simeq -2 \left| \mu(\MEW) \right|^2 -2 m_{H_u}^2(\MEW), 
\label{EWSB}
\end{eqnarray}      
where $\mu(\MEW)$ and $m_{H_u}(\MEW)$ 
are a supersymmetric higgsino mass parameter 
and a soft SUSY breaking mass for the up-type Higgs boson at the Z boson mass ($m_Z \simeq 91.2$ GeV) scale, respectively. 
The soft SUSY breaking mass $m_{H_u}$
relates to the masses of the sparticles
through the renormalization groups (RG), so $m_{H_u}(\MEW)$ tends to be large if sparticles are heavy.
Then the severer tuning between $\mu(\MEW)$ and $m_{H_u}(\MEW)$ is generally required 
to realize the observed $Z$ boson mass.
In order to derive the EW scale naturally, $m_{H_u}(\MEW)$ should be close to the EW scale as much as possible
without any conflicts with the observed Higgs mass.

We have proposed a solution to resolve this problem by taking suitable ratios among gaugino masses 
at the gauge coupling unification scale (GUT scale)~\cite{Abe:2007kf,Abe:2012xm}.
In the RG correction to $m_{H_u}(\MEW)$, there are large contributions from the top squarks
dominated by the gluino mass. 
The gluino mass easily makes the magnitude of $m_{H_u}(\MEW)$ considerably larger than the EW scale.
In the non-universal gaugino mass (NUGM) scenario, 
the RG correction from the gluino mass are canceled by those similarly from the bino and wino ones, 
thus the $m_{H_u}(\MEW)$ could be small, even if the top squark masses are relatively high.
It is remarkable that such a situation could increase the lightest Higgs boson mass 
due to the larger top squark left-right mixing, 
which is compatible with the non-universal gaugino masses for reducing the fine-tuning. 

In the NUGM scenario, the relatively heavy bino and wino are preferred: 
particularly, a ratio of wino to gluino mass, $r_2 \equiv M_2(M_{\rm GUT}) / M_3(M_{\rm GUT})$, 
should be in the range, $4 \lesssim r_2 \lesssim 6$.
Such a heavy wino enhances the left-handed top squark mass through the RG running, 
while the right-handed sparticles do not receive contributions from the wino mass.
Moreover, only the right-handed top squark becomes relatively light due to the RG effect
through the top Yukawa coupling
\footnote{The right-handed bottom squark may also be light depending on $\tan\beta$.
If $\tan\beta$ is large, the negative contributions of the bottom Yukawa coupling to the bottom squark masses are sizable.
We take $\tan \beta = 15$ 
that enhances the tree-level SM-like Higgs boson mass, 
but small enough to neglect the bottom Yukawa coupling in this paper.}.  

Aims of this paper are to show some specific features of light sparticles in the NUGM scenario 
and to investigate current experimental bounds on them at the LHC8 
and their expected sensitivities at the LHC14, especially focusing on the lightest top squark searches.
Since the naturalness argument requires the small $\mu$-parameter, 
the lightest supersymmetric particle (LSP) is a higgsino-like neutralino.
The second lightest higgsino-like neutralino and the lightest chargino are also light, 
and their mass differences are very small. 
Therefore a production cross section of the lightest top squark is sizable, 
and then top squark decays into both higgsino-like neutralinos and chargino 
together with top or bottom quarks.

The higgsino search may be also relevant to this low-scale SUSY scenario,
although their signals of the chargino decay may be buried under the Standard Model (SM) background due to their nearly degenerate masses.
There are experimental searches for the charginos and neutralinos, e.g. Refs.~\cite{Aad:2014vma,Aad:2014nua,Khachatryan:2014qwa}.
In Refs.~\cite{Baer:2013xua,Baer:2014kya,Baer:2013yha}, the way to search for the higgsino pair production at the LHC has been discussed, where small $\mu$-parameter is achieved by tuning the Higgs soft scalar masses~\cite{Baer:2012cf}. 
The higgsino search, however, would not be effective to probe the NUGM scenario.
Daughter particles of the heavier higgsino-like states will be too soft to be reconstructed in the detector, 
since the mass gaps among higgsino-like states are highly suppressed due to the heavy wino and bino. 
A strategy to search for the degenerate higgsinos is studied in Ref.~\cite{Han:2013usa}.
Note that lifetimes of the higgsinos are not so long that they are observed as disappearing tracks unlike the winos  
as explored in Refs.~\cite{Aad:2013yna,CMS:2014gxa}\footnote{
The mass difference between the lightest chargino and neutralino should be $O(0.1\,{\rm GeV})$ 
for enough large lifetime of chargino to be observed as disappearing tracks, 
while the mass difference is $O(1\, {\rm GeV})$ for higgsino-like states.}.
Moreover, heavier neutralino $\widetilde{\chi}^0_{3,4}$ and chargino $\widetilde{\chi}_2^{\pm}$ 
are so heavy that hardly produced even at the LHC14 when the gluino is heavy enough to satisfy its lower bound.
Therefore direct neutralino or chargino search will not be efficient to probe the NUGM scenario.

This paper is organized as follows. 
In Section~2, we review the NUGM scenario and explain how much the fine-tuning is relaxed while the Higgs boson mass is around 125 GeV in this scenario. 
In Section~3, specific features of the NUGM scenario is explained.
In Section~4, we discuss the current and expected experimental bounds on the scenario.
Finally, we conclude this paper in Section~5.

\section{Brief review of NUGM scenario}

We review the NUGM scenario based on Refs.~\cite{Abe:2007kf,Abe:2012xm}. 
The most attractive feature of the NUGM scenario is 
that a certain type of the non-universal gaugino mass spectrum at the GUT scale 
can relax the fine-tuning of the $\mu$-parameter 
and helps to enhance the Higgs boson mass.

The lightest CP-even Higgs boson mass in the MSSM can be  approximately written as follows at the one-loop level~\cite{Carena:1995wu}: 
\begin{eqnarray} 
m_h^2 \simeq 
m_Z^2 \cos^2 2 \beta+\frac{3}{8 \pi^2} \frac{m_t^4}{v^2} \left[\log{\frac{M_{\rm st}^2}{m_t^2}}+ 
 \frac{2\widetilde{A}_t^2}{M_{\rm st}^2} 
\left( 1-\frac{\widetilde{A}_t^2}{12 M_{\rm st}^2} \right)  \right], 
\label{Mhig}
\end{eqnarray}
where $\widetilde{A}_t \equiv A_t(\MEW)-\mu(\MEW) \cot \beta$ is defined.
$\tan\beta$ denotes the ratio of vacuum expectation values (VEVs) of two Higgs bosons: $\tan\beta \equiv \vev{H^0_u}/\vev{H^0_d}$.
The symbols $m_t$, $M_{\rm st}\equiv \sqrt{|m_{Q_3}m_{u_3}|}$ and $A_t$ denote 
the top quark mass, the top squark mass scale and the left-right mixing of the top squarks, called  A-term, respectively, 
while $m_{u_3}$ and $m_{Q_3}$ are the soft scalar masses for the right-handed top squark and the left-handed  third-generation squark, respectively. 
The Higgs boson mass is far from the observed value 
when $M_{\rm st} \lesssim 1$ TeV and the last term is negligibly small. 
The last term is maximized at $\widetilde{A}_t \sim \sqrt{6} M_{\rm st}$. 
Thus the relatively large A-term of top squarks is necessary to explain the 125 GeV Higgs boson mass.

The parameters relevant to the Higgs boson mass and the stationary conditions for the EW symmetry breaking are $m_{H_u}$, $m_{u_3}$, $m_{Q_3}$ and $ A_t$. 
Their values at the EW scale depend on the boundary condition at the GUT scale as follows: 
\begin{align}   
m_{H_u}^2(\MEW) &\simeq - 0.01 M_1 M_2 + 0.17 M_2^2 - 0.05 M_1 M_3 - 0.20 M_2 M_3 - 3.09 M_3^2 \notag \\
&+ (0.02 M_1 + 0.06 M_2 + 0.27 M_3 - 0.07 A_t) A_t + 0.59m_{H_u}^2 - 0.41m_{Q_3}^2 - 0.41m_{U_3}^2, \label{RGE_mhu}  \\
m_{Q_3}^2(\MEW) &\simeq -0.02 M_1^2+0.38 M_2^2-0.02 M_1 M_3-0.07 M_2 M_3+5.63 M_3^2 \notag \\ 
&+(0.02 M_2+0.09 M_3-0.02 A_t)A_t-0.14m_{H_u}^2+0.86m_{Q_3}^2-0.14 m_{U_3}^2, \label{RGE_Q} \\ 
m_{u_3}^2(\MEW) &\simeq 0.07 M_1^2-0.01 M_1 M_2-0.21 M_2^2-0.03 M_1 M_3-0.14 M_2 M_3+4.61 M_3^2 \notag \\ 
&+(0.01M_1+0.04 M_2+0.18 M_3-0.05 A_t) A_t-0.27 m_{H_u}^2-0.27 m_{Q_3}^2+0.73m_{U_3}^2, \label{RGE_u} \\
A_t(\MEW) &\simeq -0.04 M_1-0.21 M_2-1.90 M_3+0.18 A_t, \label{RGE_At}
\end{align} 
where $M_i\ (i=1,2,3)$ are the gaugino mass parameters.
Note that all the parameters in the right-hand side are those evaluated at the GUT scale. 

In Eq. (\ref{RGE_mhu}), we can see that the absolute value of $m_{H_u}(\MEW)$ becomes large, 
when the value of $m_{u_3},\ m_{Q_3}$ or $A_t$ increases.
Moreover, Eqs.(\ref{RGE_Q})-(\ref{RGE_At}) tell us that the ratio $A_t(\MEW) / M_{\rm st}$ cannot be so large 
as far as the gluino mass $M_3$ dominates the RG corrections. 
Thus the naturalness and the 125 GeV Higgs boson mass are hard to be achieved simultaneously.

Such a situation can be avoided by considering large values of wino and bino masses, 
particularly wino mass gives more significant contributions than the bino mass does. 
The contribution of the gluino mass ($M_3$) to the RG running of $m_{H_u}(\MEW)$ can be canceled 
by those of the wino (and sub-dominantly bino) masses ($M_2$ and $M_1$) as can be read off in Eq.(\ref{RGE_mhu}).
More precisely,  $m_{H_u}(\MEW)$ can remain small although top squark masses increase 
if the wino to gluino mass ratio at the GUT scale ($M_2/M_3 \equiv r_2$) is in the range $4 \lesssim r_2 \lesssim 6$~\cite{Abe:2012xm}.

The suitable ratio of wino to gluino mass does not only relax the degree of tuning, 
but also enhances the Higgs boson mass due to a relatively large value of the A-term.
In Eqs.(\ref{RGE_Q}) and (\ref{RGE_u}), 
we see that only the left-handed third-generation squark mass $m_{Q_3}(\MEW)$ increases 
when the wino mass increases, while the right-handed top squark mass $m_{u_3}(\MEW)$ does not. 
Furthermore, $m_{u_3}(\MEW)$ tend to be small according to the top Yukawa coupling, compared with
the other sparticle masses. 
As a result, only the right-handed top squark is significantly lighter than the other sfermions.
Besides, the absolute value of $A_t(\MEW)$ will become large when the wino (bino) mass increases, 
since gaugino masses contribute to RG runnings of A-terms with the same sign.
Consequently, a large ratio of A-term to top squark mass scale $A_t(\MEW) / M_{\rm st}$ can be accomplished, 
and then the Higgs boson mass can reach 125 GeV even when the top squark mass is lighter than 1 TeV.
Note that we assume the gaugino masses dominate the RG-runnings, 
in other words, A-terms and scalar masses are not extremely larger than the gaugino masses. 

\section{Phenomenological features of NUGM scenario}
In our analysis, we assume the soft SUSY-breaking scalar masses, including the Higgs soft masses, are universal
and A-terms are flavor-independent just for simplicity, 
because these parameters do not play essential roles in the NUGM scenario. 
The values at the GUT scale are denoted by $m_0$ and $A_0$ respectively. 
Thus there are six independent parameters in our analysis:
three gaugino masses $M_1, M_2, M_3$, universal scalar mass $m_0$, flavor independent A-term $A_0$.

For simplicity, we fix $\tan\beta = 15$ in our analysis,
with which the contributions of the bottom Yukawa couplings are negligibly small.
Furthermore, $m_0$ is fixed at $1$ TeV, because our main interest is the gaugino mass dependence. 
In Sec. \ref{sec4}, $M_2$ and $A_0$ are tuned to realize $\mu = 150$ GeV and $125\le m_h < 126$ GeV,
and the parameter space of $(M_1, M_3)$ is scanned in our analysis. 
We use the softsusy-3.5.1~\cite{Allanach:2001kg} to evaluate masses and mixings of sparticles and Higgs bosons.

\begin{figure}
\centering
\hfill
\includegraphics[width=0.45\linewidth]{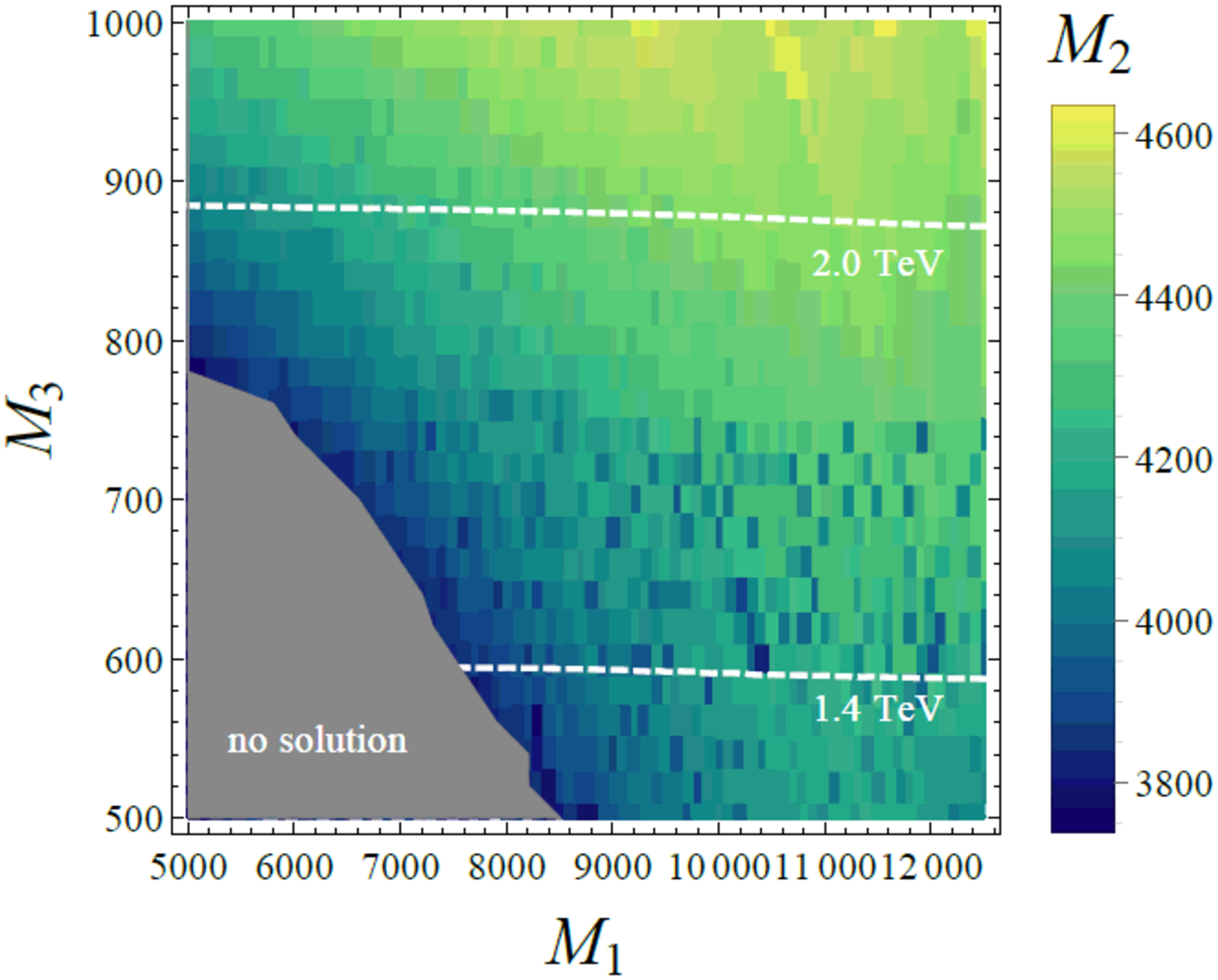} 
\hfill 
\includegraphics[width=0.45\linewidth]{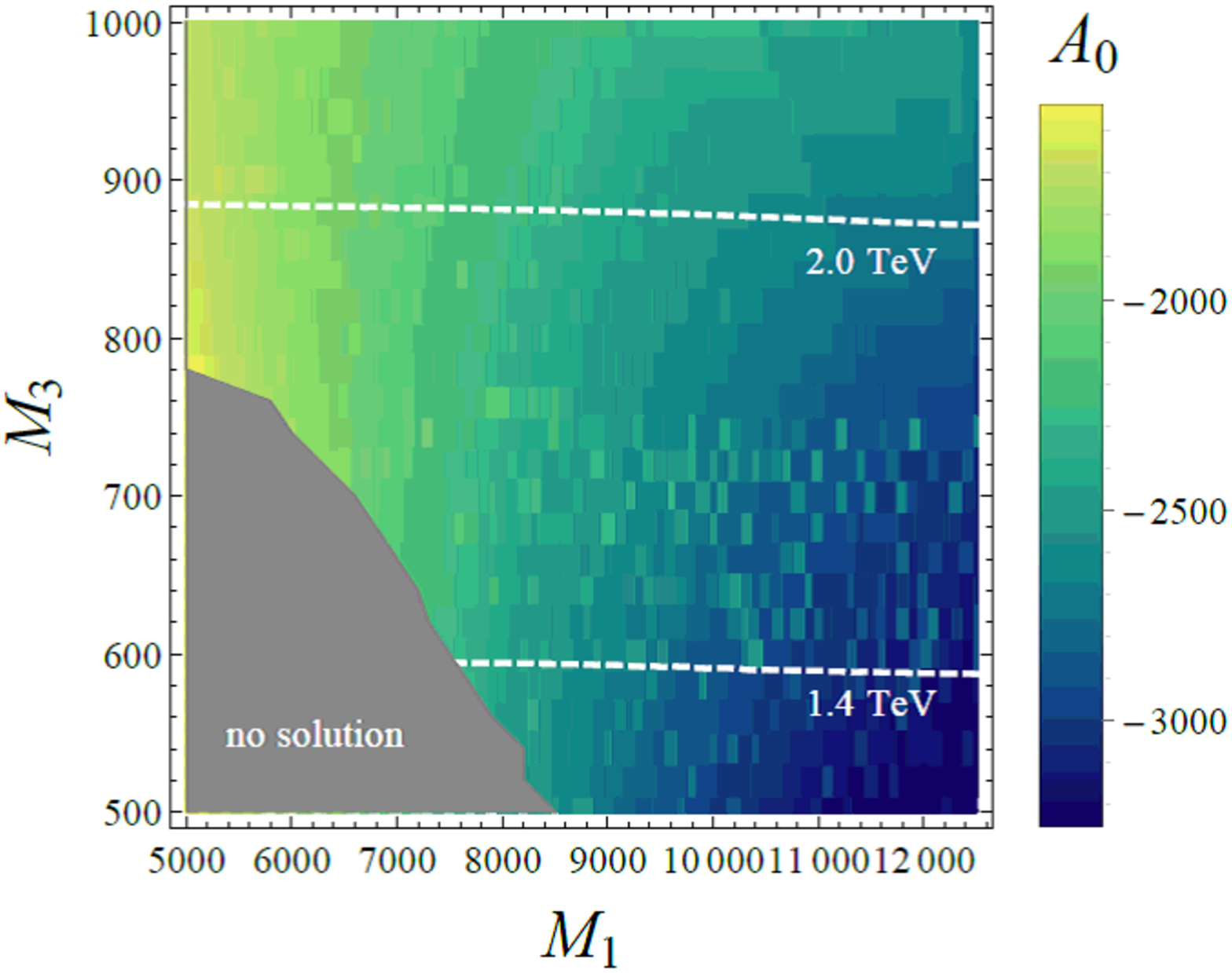}
\hfill 
\caption{Values of input parameters, $M_2$ (left panel) and $A_0$ (right panel), at the GUT scale 
that achieve $\mu = 150$ GeV and $125\le m_h < 126$ GeV.
The white dashed lines represent the mass of the gluino at the EW scale.
We cannot find input parameters that satisfy our requirements in the gray region.
The values of $M_1$, $M_2$, $M_3$ and $A_0$ are shown in the unit GeV.}
\label{fig_input}
\end{figure}

First, let us discuss $M_2$ and $A_0$ for $\mu = 150$ GeV and $125\le m_h < 126$ GeV.
Fig.\ref{fig_input} shows the sizes of $M_2$ (left panel) and $A_0$ (right panel) on the plane of $(M_1, M_3)$. 
We can see that typical values of $M_2$ and $A_0$ are around 4.0 TeV and $-2.5$ TeV.
In the gray region, 
the lightest top squark becomes tachyonic, so $M_1$ and $M_3$ cannot be so small to achieve the small $\mu$ term.
{

\begin{table}
\centering
\begin{tabular}{|c|c|c|c|}\hline
  input [GeV]& sample 1 & sample 2 & sample 3 \\ \hline\hline
 $\mu(\MEW)$ & 150 & 150 & 150 \\
 $m_0$ & 1000 & 1000 & 1000 \\
 $A_0$ & -1950 & -2400 & -2500 \\
 $M_1$ & 6500  &  9000 & 10000 \\
 $M_2$ & 4231  &  4458 & 4478  \\
 $M_3$ & 900  &  900 & 900  \\ \hline\hline
 mass [GeV] & & & \\ \hline\hline
 $m_{\widetilde{t}_1}$ &695.2 &1169 & 1414 \\
 $m_{\widetilde{b}_1}$ & 1971 & 2081 & 2139 \\
 $m_{\widetilde{g}}$ & 2035 & 2041 & 2046 \\
 $m_{\widetilde{\chi}^{\pm}_1}$ & 154.6 & 155.1 & 154.7 \\
 $m_{\widetilde{\chi}^0_1}$ & 152.9 & 153.6 & 153.3 \\
 $m_{\widetilde{\chi}^0_2}$ & 156.1 & 156.3 & 155.8 \\
 $m_{\widetilde{\chi}^0_3}$ & 2884 & 3621 & 3639 \\
 $m_{\widetilde{\chi}^0_4}$ & 3435 & 4004 & 4453 \\ \hline\hline
branching ratio & & & \\ \hline\hline
 Br$(\widetilde{t}_1 \rightarrow t \widetilde{\chi}^0_1)$ &0.228 & 0.242 & 0.245 \\ 
 Br$(\widetilde{t}_1 \rightarrow t \widetilde{\chi}^0_2)$ & 0.240 & 0.249 & 0.251 \\ 
 Br$(\widetilde{t}_1 \rightarrow b \widetilde{\chi}^{\pm}_1)$ & 0.532 & 0.509 & 0.505 \\
 Br$(\widetilde{g} \rightarrow \widetilde{t}_1 \bar{t})$ & 0.497 & 0.500 & 0.500 \\  \hline\hline
output & & & \\ \hline\hline
 $N_{\rm signal}$ & 5.04& $3.71 \times 10^{-2}$ & $6.71 \times 10^{-3}$ \\ 
 $S / \sqrt{B}$ & 43.2 & 5.41 & 1.71  \\ \hline
\end{tabular}
\caption{Values of parameters at some sample points.
$N_{\rm signal}$ is the number of events coming from top squark pair production 
with the center of mass energy $\sqrt{s} = 8$ TeV 
with an integrated luminosity 20.1 ${\rm fb}^{-1}$ in a signal region with $m_{\rm CT}>350$ GeV 
of Ref.~\cite{Aad:2013ija}. 
$S/\sqrt{B}$ is the significance of the signal events, 
where $S$ and $B$ are the number of  signal and background events respectively
with a center of mass energy $\sqrt{s} = 14$ TeV and an integrated luminosity of 300 ${\rm fb}^{-1}$. 
For the numbers of backgrounds we refer to Ref.~\cite{ATLAS:2014high}, 
and then we select a signal region that maximizes the significance for each sample point. }
\label{tab_sample}
\end{table}

Table \ref{tab_sample} shows some benchmark points.
We can see that the top squark $\widetilde{t}_1$ is significantly lighter 
than bottom squark $\widetilde{b}_1$ and gluino $\widetilde{g}$.
Since the masses of bino and wino are around or larger than the gluino mass at the EW scale, 
then sleptons tend to be as heavy as squarks, other than the top squark, 
in contrast to the most of well-known scenarios, 
e.g. the CMSSM or the minimal gauge mediation~\cite{Giudice:1998bp}.

In the NUGM, a top squark is lighter than gluino. 
Then the lower bound on the gluino mass will be given by the analysis in Ref.~\cite{Aad:2014lra} 
rather than the analysis in Ref.~\cite{Aad:2014wea}.
In the former, the exclusion limit of the gluino mass reaches $1.4$ TeV.
This bound would depend on decay properties of the top squark, 
and then the exact bound on the NUGM is unknown. 
We will study gluino searches in the NUGM elsewhere, 
while in this paper we just expect that the bound on the gluino is around the result of Ref.~\cite{Aad:2014lra}. 
Then we concentrate on the parameter region of the NUGM with $m_{\widetilde{g}} \gtrsim 1.4$ TeV
that corresponds to $M_3 \gtrsim 600$ GeV.
Note that the other sparticles could not be produced at the LHC, 
since they are heavier than the gluino.

The masses of the lightest chargino $\widetilde{\chi}^{\pm}_1$ 
and the lightest and the second lightest neutralino $\widetilde{\chi}^{0}_{1,2}$ are just above $\mu = 150$ GeV, 
since they are virtually higgsino-like.
In addition, 
the mixings among the higgsino-like states and bino- or wino-like states are highly suppressed, since $\widetilde{\chi}^{0}_{3,4}$, which are bino-like and wino-like neutralinos, are quite heavy.
As a result, the mass differences between the higgsino-like states are highly suppressed.

The degenerate higgisnos will make $\widetilde{\chi}_2^0$, $\widetilde{\chi}_1^{\pm}$ decay
into lighter higgsino-like states invisible.
The left panel of Fig.\ref{fig_mdiff_Br} shows the mass difference 
between $\widetilde{\chi}_1^0$ and $\widetilde{\chi}_1^{\pm}$.
We can see that the mass difference is less than $2$ GeV 
and then their decay products would be lower than the reconstruction threshold.
The mass difference between $\widetilde{\chi}_2^0$ and $\widetilde{\chi}_1^{\pm}$ is also less than 2 GeV.

\begin{figure}
\centering
\hfill
\includegraphics[width=0.45\linewidth]{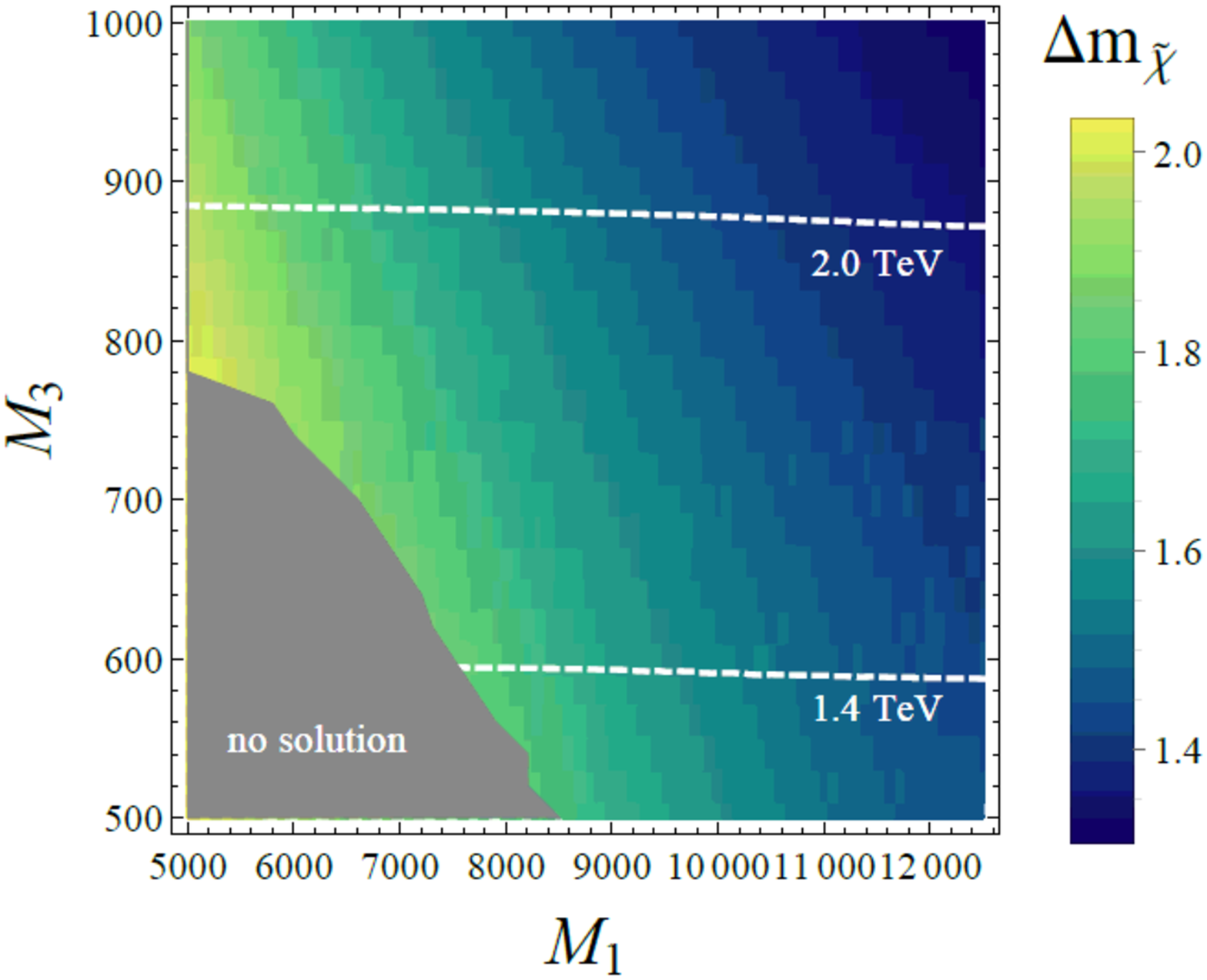}
\hfill 
\includegraphics[width=0.45\linewidth]{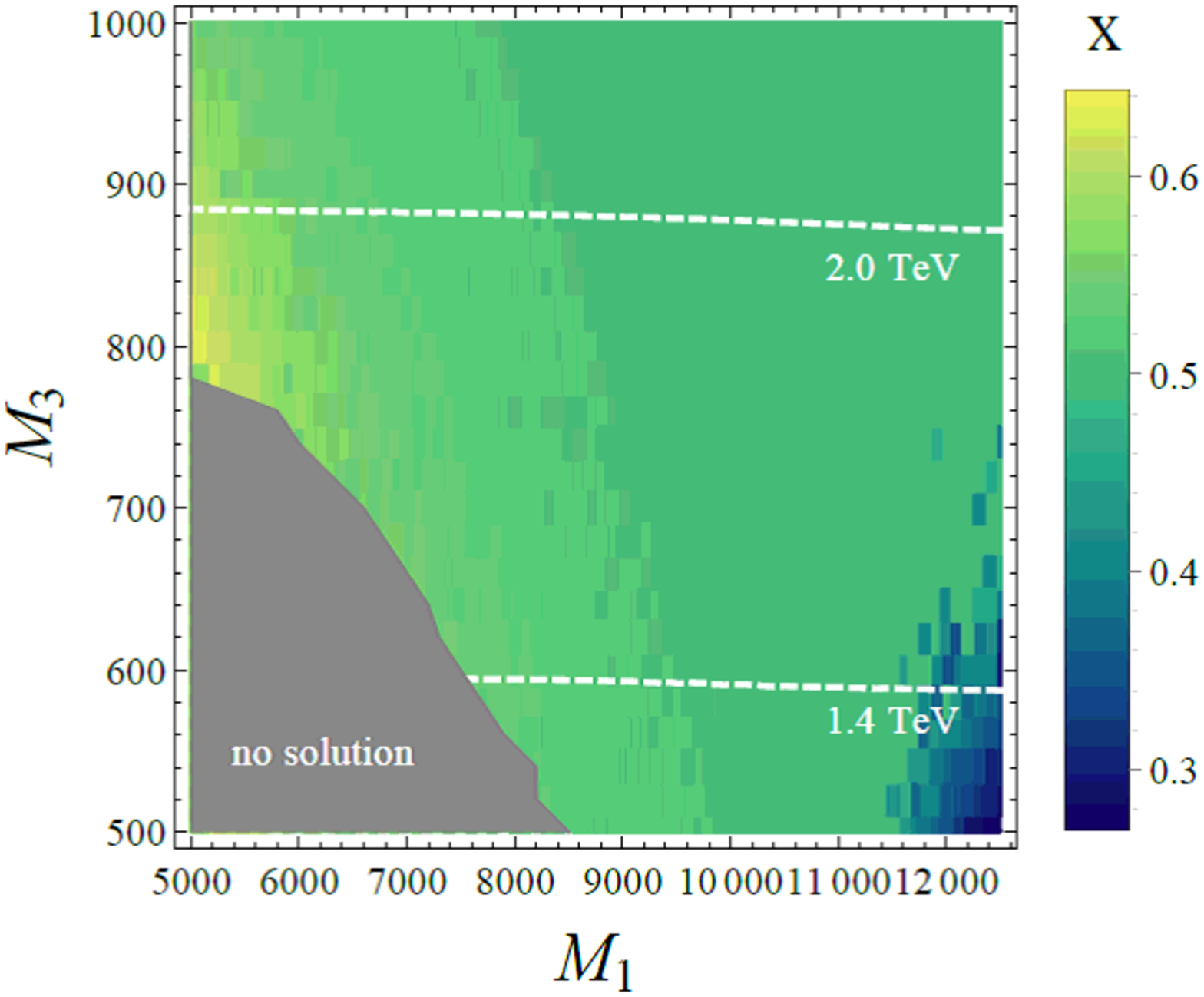} 
\hfill 
\caption{The mass difference between $\widetilde{\chi}^{\pm}_1$ and $\widetilde{\chi}^0_1$ (left panel) 
and the branching ratio of the top squark to the bottom quark and the lightest chargino (right panel),  
$X \equiv {\rm Br}(\widetilde{t}_1 \rightarrow b \widetilde{\chi}_1^{\pm})$, are shown. 
The meanings of the lines and colored regions are the same as Fig.\ref{fig_input}. 
The values of $M_1$, $M_3$ and $\Delta m_{\tilde{\chi}}$ are shown in the unit GeV. }
\label{fig_mdiff_Br}
\end{figure}

The decay channels of the top squark are almost fixed in the NUGM. 
The top squark couples to neutralino and chargino through the top Yukawa coupling, 
and the lightest state is almost right-handed.
Moreover, the masses of the higgsino-like states are nearly degenerate as explained above. 
These facts fix the branching ratios of the top squark as, 
\begin{align}
{\rm Br}(\widetilde{t}_1 \rightarrow t \widetilde{\chi}_1^0) &\simeq 25 \%, \\
{\rm Br}(\widetilde{t}_1 \rightarrow t \widetilde{\chi}_2^0) &\simeq 25 \%, \\
{\rm Br}(\widetilde{t}_1 \rightarrow b \widetilde{\chi}_1^{\pm}) &\simeq 50 \%. 
\end{align}
The branching ratios, ${\rm Br}(\widetilde{t}_1 \rightarrow b \widetilde{\chi}_1^{\pm}) \equiv X$, are shown 
in the right panel of Fig. \ref{fig_mdiff_Br}. 
We use SDECAY~\cite{Muhlleitner:2003vg} to compute the decay widths and the branching ratios of sparticles.
The numerical values of $X$ significantly decrease around $(M_1, M_3) \sim (12, 0.55)$ TeV 
where the top squark can decay into or through the gluino.

\section{Top squark search at the LHC}
\label{sec4}

Finally, we discuss the signals of our top squark at the LHC.
We use the Madgraph 5~\cite{Alwall:2014hca} to simulate the signal events at the parton-level.
After the event selections with suitable cuts, the number of events are compared 
with the 95$\%$ confidence level (C.L.) upper limits on the number of signal events 
given by the ATLAS collaboration~\cite{Aad:2013ija}. 
We refer 
to Table 11 of Ref.~\cite{ATLAS:2014high} 
for the expected numbers of the background events at the LHC14.  
The higgsino-like states $\widetilde{\chi}^{\pm}_1, \widetilde{\chi}^0_{1,2}$ are treated as invisible particles in our analysis, 
since the transverse momentum of SM particles produced by the higgsino decay would be below the reconstruction threshold. 

\subsection{Details of the analysis}

We generate $10^5$ signal events at each point with $\sqrt{s}=$ 8 TeV or 14 TeV, 
then these are normalized to be consistent with the integrated luminosity 
that is observed or expected at the LHC, respectively.

Some properties of event reconstruction procedures are taken into account in our analysis 
with respect to Ref.~\cite{Aad:2013ija}. 
Electrons (muons) must have a transverse momentum, denoted as $p_T$, larger than 7(6) GeV 
and their pseudo-rapidity must be in a range $|\eta| < 2.47(2.4)$, otherwise leptons are discarded. 
Here a pseudo-rapidity  is defined as $\eta \equiv -\log( \tan \theta / 2)$, 
where $\cos \theta \equiv p_z / |\bm p|$.\footnote{
The z-axis is along the incident beam direction and $\bm p$ is a spatial momentum vector}
If quarks except top quarks have $p_T > 20$ GeV and $|\eta| < 4.9$, 
they are counted as jets. 
Then, if its flavor is bottom and $|\eta| < 2.5$, 
it is interpreted as b-tagged jet with a b-tagging efficiency. 
We assume the b-tagging efficiency is 60$\%$ in our analysis. 
Finally, a missing transverse momentum ${\bm p}_{T}^{\rm miss}$ of each event is defined 
as the one opposite to a sum of all visible particles within $|\eta| < 4.9$, 
and then missing transverse energy is defined as $E_T^{\rm miss} \equiv |{\bm p}_T^{\rm miss}|$.

If some of jets or leptons are overlapped in a $(\phi, \eta)$ plane, where $\phi$ is an azimuthal angle, 
they are resolved following the experimental analyses.
The transverse momentum of two jets are summed and $\eta, \phi$ are summed 
weighted by each transverse momentum 
when the distance between two jets is within $\Delta R < 0.4$, 
where $\Delta R \equiv \sqrt{(\Delta \eta)^2 + (\Delta \phi)^2}$ is defined.
If one of two jets is b-tagged, the jet after the resolving procedure is treated as a b-tagged jet. 
Overlaps between light-flavor jets and electrons within $\Delta R < 0.2$ are resolved 
by discarding the jet, while the electron is discarded if the jet is b-tagged. 
When the overlaps between electrons (muons) and any jet are within $0.2 < \Delta R < 0.4$ ($\Delta R < 0.4$), both of the electron and the muon are discarded.

\subsection{Results of the analysis}
Direct top squark searches have been done dedicated to the several channels 
at the ATLAS~\cite{Aad:2013ija,Aad:2014bva,Aad:2014kra,Aad:2014qaa} 
and the CMS~\cite{Khachatryan:2015wza,Khachatryan:2015vra,Khachatryan:2015pwa,Khachatryan:2015lwa,
Chatrchyan:2013xna}.

Pair-produced top squarks decay into the SM particles in several ways.
When a top squark decays into a top quark and a neutralino, $\tilde{t}_1\rightarrow t \tilde{\chi}_{1,2}^0$, 
the top quark decays into a bottom quark and a W boson 
which subsequently decays into two light-flavor quarks, or a charged lepton and a neutrino.
Then the pair-produced top squarks decay as
$\tilde{t}_1 \tilde{t}^*_1 \rightarrow t \bar{t} +  \tilde{\chi}^0\tilde{\chi}^0 
\rightarrow b \bar{b} + f \bar{f} f' \bar{f'} + \tilde{\chi}^0 \tilde{\chi}^0$, 
where $f \bar{f}$ and $f' \bar{f'}$ are $q \bar{q}$ or $l \nu$. 
Thus corresponding channels are $2b + (4j\ {\rm or}\ 2j + 1l\ {\rm or}\ 2l) + E_T^{\rm miss}$.

The $2b+4j+E_T^{\rm miss}$ and $2b + 2j + 1l + E_T^{\rm miss}$ channels 
give more stringent bound on the top squark mass than the $2b + 2l + E_T^{\rm miss}$ channel, 
since the hadronic decay of W boson is about 70 $\%$
although the leptonic channels would not be suffered from the SM background.
In the $2b+4j+E_T^{\rm miss}$ channel,
one of the dominant SM backgrounds is the $t\bar{t}$ production, 
where one top quark decay semileptonically and the lepton is not reconstructed successfully,  
and the other is the $Z(\rightarrow \nu\nu)+{\rm jets}$~\cite{Aad:2014bva}. 
In the $2b + 2j + 1l + E_T^{\rm miss}$ channel, 
the SM backgrounds are dominated by the $t\bar{t}$ production,  
where the both top quarks decays semileptonically and one lepton is not reconstructed, 
and the $W +{\rm jets}$ where the W boson decays leptonically~\cite{Aad:2014kra}.

When a top squark decays into a bottom quark and a chargino, $\tilde{t}_1 \rightarrow b \tilde{\chi}^{\pm}_1$, 
the chargino decays into the lightest neutralino and two light-flavor quarks or a charged lepton and a neutrino, 
where the two SM fermions are produced through the off-shell W boson 
if the mass difference between the neutralino and the chargino is less than the W boson mass.
Then the pair-produced top squarks decay as, 
$\tilde{t}_1 \tilde{t}^*_1 \rightarrow b \bar{b} +  \tilde{\chi}^{+}_1\tilde{\chi}^{-}_1 
\rightarrow b \bar{b} + f \bar{f} f' \bar{f'} + \tilde{\chi}^0 \tilde{\chi}^0$.
The signal is similar to the case of 
$\tilde{t}_1 \tilde{t}^*_1 \rightarrow t \bar{t} +  \tilde{\chi}^0\tilde{\chi}^0$.
However the daughter particles of the chargino become soft
due to the small mass difference between the neutralino and the chargino, 
and then soft leptons are required by the event selections 
in many analyses targeting to the top squarks decaying into bottom quarks and charginos.

Such soft daughter particles of the chargino could be too soft to be reconstructed at the detector 
if the mass difference between the neutralino and the chargino is quite small. 
This occurs exactly in the NUGM scenario, since their mass difference is smaller than 2 GeV. 
Thus the corresponding channel to this case is $2b + E_T^{\rm miss}$ 
where all of the daughter particles of the chargino are not reconstructed. 
In this channel, the dominant SM background is $Z(\rightarrow\nu\nu) + {\rm jets}$ 
where jets should include b-jets, 
and the backgrounds coming from top quarks will become sub-dominant~\cite{Aad:2013ija}.

Since the branching ratio of $\widetilde{t}_1 \rightarrow t \widetilde{\chi}^0_{1,2}$ is around 50 $\%$, 
the analyses dedicated to the channels $t\bar{t} + E_T^{\rm miss}$ may not be so efficient in the NUGM scenario. 
The corresponding number of signal events 
is quarter, compared with the case with Br$(\widetilde{t}_1 \rightarrow t \widetilde{\chi}^0_{1,2}) = 100\%$. 
This is almost same as the number of signal events in the $2b+E_T^{\rm miss}$ channel. 
Thus the most efficient channel to investigate the NUGM scenario would be $2b + E_T^{\rm miss}$, 
because the SM background from $Z(\rightarrow\nu\nu) + {\rm jets}$ 
would be easier to be distinguished from the signal events than those of $t\bar{t}$ production.  
These arguments will depend on details of analyses, 
but these observations are also pointed out and confirmed in Ref.~\cite{Han:2013kga}.


The 95$\%$ C.L. upper limit on the number of the SUSY event is displayed in Table 7 of Ref.~\cite{Aad:2013ija} 
that is analyses based on the data with $\sqrt{s} = 8$ TeV and the integrated luminosity of $20.1\ {\rm fb}^{-1}$. 
The number of signal events after the same event selection criteria is calculated following the method explained 
in the previous subsection.
The number of survived events at the sample points are shown in Table \ref{tab_sample}.

The expected number of the background events dedicated to $2b + E_T^{\rm miss}$ 
with $\sqrt{s} = 14$ TeV and with an integrated luminosity of $300\ {\rm fb}^{-1}$ is studied 
in Ref.~\cite{ATLAS:2014high}. {
The number of events can be seen in Table 11 of Ref.~\cite{ATLAS:2014high}.
In fact, this analysis is devoted to the search for the bottom squark, through the
decay as $\widetilde{b}_1 \rightarrow b \widetilde{\chi}^0_1$. 
However, the expected final states and their kinematics are quite similar to 
$\widetilde{t}_1\rightarrow b \widetilde{\chi}^{\pm}_1$,
if the chargino $\widetilde{\chi}_1^{\pm}$ can be treated as the invisible particle effectively.
The expected significance $Z \equiv S/\sqrt{B}$, 
where $S$ is the number of signal events after the event selection following the analysis 
in Ref.\cite{ATLAS:2014high}, 
is shown in the Table \ref{tab_sample}.

\begin{figure}
\centering
\includegraphics[width=0.65\linewidth]{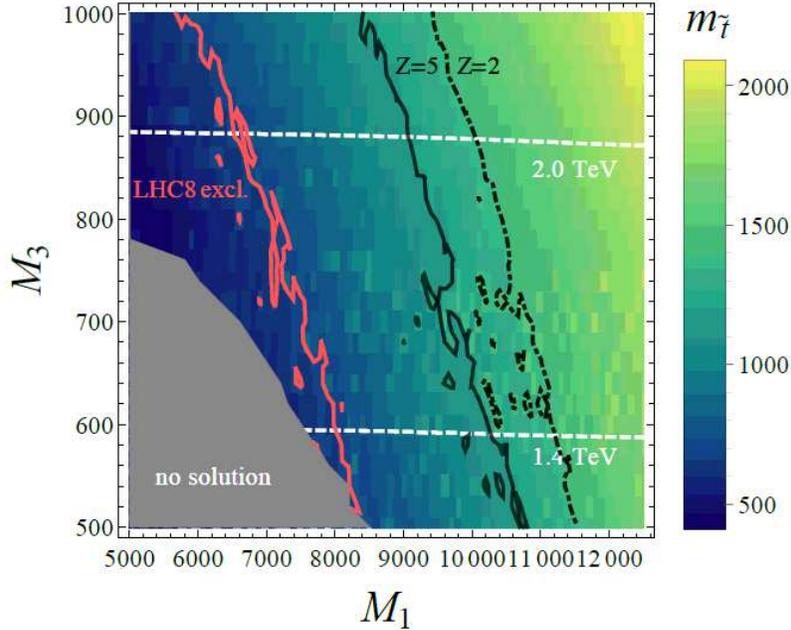}
\caption{Masses of the top squark and its experimental bound. 
The red line is the 95$\%$ C.L. limits from the data 
with $\sqrt{s} = 8$ TeV and an integrated luminosity of 20.1 ${\rm fb}^{-1}$. 
The black lines represent the expected significance $Z \equiv S / \sqrt{B}$ 
with $\sqrt{s} = 14$ TeV and an integrated luminosity of 300 ${\rm fb}^{-1}$.
The meanings of the white lines and colored regions are the same as Fig. 1. The values of $M_1$, $M_3$ and $m_{\tilde{t}}$ are shown in the unit GeV. }
\label{fig_stop}
\end{figure}

Fig.\ref{fig_stop} shows the top squark mass and the experimental limits obtained from our analyses.
In this figure, the red line describes the 95$\%$ C.L. exclusion upper limit obtained 
from the data with $\sqrt{s} = 8$ TeV
and the black lines represent the significance $Z \equiv S/\sqrt{B} = 2\ {\rm (dashed)}\ {\rm and}\ 5\  {\rm (solid)}$.
There are several signal regions corresponding to different cut 
with respect to the values of the contransverse mass $m_{\rm CT}$.
The most sensitive signal region is selected at each point.

We can see that the current experimental lower bounds on the top squark mass in the NUGM is about $690$ GeV.
The bounds on the gaugino masses are $M_1 \gtrsim 6.0-8.0$ TeV depending on the value of $M_3$.
The top squark mass is mostly determined by $M_1$ in our analysis.
Consequently the top squark mass is sensitive to $M_3$ in general, 
but its contribution is subtracted by those of $M_2$ in the NUGM scenario
according to the constraint on the small $\mu$ term, or equivalently small $|m_{H_u}|$. 
The top squark mass is sensitive to $M_1$, while $m_{H_u}$ is relatively insensitive to it~\cite{Abe:2007kf}.


Fig.\ref{fig_stop} also shows the expected  sensitivity at the LHC14 
with an integrated luminosity of $300\ {\rm fb}^{-1}$. 
It is found that the significance reaches $Z = 5$ that maybe correspond to the discovery for 
$ m_{\widetilde{t}_1}  \simeq1.2$ TeV, 
and reaches $Z=2$ that may correspond to an exclusion limit for $m_{\widetilde{t}_1}  \simeq 1.4$ TeV. 

\section{Conclusion} 
In this paper, we have studied phenomenological properties relevant to collider experiments
of the non-universal gaugino masses (NUGM) scenario.
Especially, we have discussed the current and expected sensitivities of the top squark searches at the LHC.

The NUGM scenario can relax the fine-tuning of the $\mu$-parameter~\cite{Abe:2007kf}, 
while can enhance the SM-like Higgs boson mass through the RG-runnings~\cite{Abe:2012xm}. 
The key ingredient of the NUGM is a ratio of the wino to gluino mass, $r_2 \equiv M_2/M_3$, 
at the GUT scale which should be in a range $4 \lesssim r_2 \lesssim 6$~\cite{Abe:2007kf,Abe:2012xm}.
In this case, the lightest top squark, which is almost right-handed one, 
is lighter than all the other sfermions and gauginos.

The gaugino mass ratios should be fixed precisely in order to avoid the fine-tuning, 
then we have to pay attention to how the desirable ratios are realized.
The NUGM scenario can be obtained from some UV physics.
An interesting possibility is the TeV-scale mirage mediation~\cite{Choi:2005uz,Endo:2005uy,Choi:2006xb}, 
that is, a mixed SUSY breaking mediation 
via the moduli and the conformal anomaly~\cite{Randall:1998uk}. 
The desired cancellation among the gaugino masses in RG runnings occurs, 
even when the so-called mirage unification relation does not hold exactly, 
or there are gauge mediated contributions in addition to the mirage mediation~\cite{Everett:2008qy} 
if it has suitable values~\cite{Abe:2014kla}.
The gaugino masses are controlled by the ratio of the contributions from moduli and anomaly mediation, 
which depends on the moduli stabilization scenarios~\cite{Abe:2005rx}.

The non-universal gaugino masses can also be realized even at the tree-level 
(in four-dimensional effective field theory).
For example, 
certain linear-combinations of multiple muduli fields appear in the gauge kinetic functions
in some superstring models with nontrivial D-brane configurations, 
and then gaugino mass ratios are determined by 
e.g. the numbers of winding, intersection or magnetic fluxes 
for D-branes from which the SM gauge bosons arises~\cite{Blumenhagen:2006ci}.
In the case that the gauge boson for each SM gauge group lives on a different D-brane 
from each other, their gauge kinetic functions are different linear-combinations, 
that could be the origin of gaugino mass non-universality. 
Furthermore, the $U(1)_Y$ symmetry of the SM may be given by a linear combination of multiple $U(1)$ symmetries with quite different origins. This yields a correction to the bino mass of its own. 
Finally, even in the four-dimensional SUSY GUT framework, 
SUSY breaking fields contained in the gauge kinetic function may not be singlet under the GUT symmetry.
They will give non-universal masses to gauginos depending on their representations~\cite{Younkin:2012ui}.



Let us comment on dark matter candidates in the NUGM scenario.
The LSP is typically purely higgsino-like and its mass is $\mathcal{O}(100{\rm GeV})$. 
This means the amount of thermally produced LSP 
is quite less than the cosmologically required value~\cite{Mizuta:1992qp}. 
The relic density can be accomplished 
when new dark matter candidate is introduced in addition to the LSP, such as axions, 
and/or there is an enough amount of non-thermally produced LSP comes from decays of long-lived heavy particles 
such as gravitino~\cite{Kohri:2005ru} and/or modulus fields~\cite{Moroi:1999zb}.
The collider physics would be unchanged in these cases.  
Even when we consider thermally produced dark matter composed of only the LSP, 
the NUGM scenario can also provide the suitable dark matter candidate.
One way is that LSP is mixture of higgsino and bino 
that can be achieved in the parameter region $M_1 \ll M_2, M_3$ 
although such region is out of the figures in this paper~\cite{ArkaniHamed:2006mb}.
Another possibility is to introduce new sparticle lighter than the neutralinos 
in extended models of the MSSM.
Interesting candidates are an axino which is a superpartner of axion~\cite{Rajagopal:1990yx}, 
or the singlino in the Next-to-MSSM~\cite{Ellwanger:2009dp}. 
In these cases, the collider phenomenology would be altered from the analyses in this paper.

Since the mass differences 
between the lightest neutralino and the lightest chargino $\Delta m_{\widetilde{\chi}^{\pm}_1 - \widetilde{\chi}^0_1}$, 
or the second-lightest neutralino and the lightest chargino $\Delta m_{\widetilde{\chi}^{0}_2 - \widetilde{\chi}^{\pm}_1}$ 
are typically less than 2 GeV, 
all of decay products of the heavier higgsino-like states are invisible.
The lightest top squark decays into $t \widetilde{\chi}^0$ and $b \widetilde{\chi}^{\pm}$ 
with almost the same branching ratio, 
because the top squark is made of mostly the right-handed one. 
These features make it difficult to search for top squark production 
dedicated to $t\bar{t} + E_T^{\rm miss}$ or $b\bar{b} + f\bar{f}f'\bar{f'} + E_T^{\rm miss}$. 
Thus the $b\bar{b} + E_T^{\rm miss}$ channel gives the most stringent bound on the top squark mass.

We find out the experimental 95$\%$ C.L. exclusion limit on the NUGM 
by referring to the result of the analysis dedicated to $b\bar{b} + E_T^{\rm miss}$ channel 
with $\sqrt{s} = 8$ TeV data and an integrated luminosity of $20.1\ {\rm fb}^{-1}$. 
The lower bound on the top squark mass is about 690 GeV 
and the allowed region on the $(M_1, M_3)$ plane can be shown in Fig.\ref{fig_stop}.

We also studied the expected significance at the LHC with $\sqrt{s} = 14$ TeV and $300$ ${\rm fb}^{-1}$.
The significance of the signal of the $b\bar{b} + E_T^{\rm miss}$ channel will reach $Z = 5$ 
if the lightest top squark mass is less than $1.2$ TeV 
and will reach $Z = 2$ if the mass is less than $1.4$ TeV as can be seen in Fig.\ref{fig_stop}.
Thus the top squark lighter than a mass scale around $1.2$ TeV will be discovered 
and the top squark lighter than about $1.4$ TeV could be excluded. 

Finally, let us comment on the other possibilities to probe the NUGM scenario. 
Firstly, a single top quark channel, $tb + E_T^{\rm miss}$, would be promising to probe the NUGM, 
since the half of the decay from the top squark corresponds to the signal, $tb + E_T^{\rm miss}$, 
due to the branching ratio of the top squark, ${\rm Br}(\widetilde{t}_1 \rightarrow t \widetilde{\chi}_{1,2}^0) 
    \simeq {\rm Br}(\widetilde{t}_1 \rightarrow b \widetilde{\chi}_1^{\pm}) \simeq 50 \%$ .
Thus this channel could give the same or even severer limits on the top squark masses.
The top squark becomes light in the small $M_1$ region, 
while the gluino search will become the most important in the small $M_3$ (and large $M_1$) region. 
Since the gluino will decay into the top squark in the latter region,  
then the features of the top squark decays showed in this paper 
will also be important for the gluino searches.
We can also consider the case where $\tan\beta$ is so large that the bottom Yukawa coupling 
becomes the same order as the top one.
In such case, the right-handed bottom squark would be lighter than the other sfermions, 
then the bottom squark becomes also accessible at the LHC.
We will study these possibilities in the future.


 \afterpage{\clearpage}

\subsection*{Acknowledgements}
This work was supported in part by the Grant-in-Aid for Scientific Research No. 25800158 (H.A.) and No. 23104011 (Y.O.) from the Ministry of Education, 
Culture, Sports, Science and Technology (MEXT) in Japan. 

\appendix

\end{document}